\documentclass[a4paper,11pt]{article}
\pdfoutput=1 

\usepackage{jcappub} 

\usepackage[T1]{fontenc} 

\usepackage{graphicx}
\usepackage{xcolor}
\usepackage{diagbox}

\newcommand*{\email}[1]{%
    \normalsize\href{mailto:#1}{#1}\par
    }

\begin{document}
\title{Impact of dynamical dark energy on the neutron star equilibrium}
\author{S. Smerechynskyi$^1$}
\email{sviatoslav.smerechynskyi@lnu.edu.ua}
\author{M. Tsizh$^1$}
\author{B. Novosyadlyj$^{2,1}$}
\affiliation{$^1$Ivan Franko National University of Lviv, Kyryla and Methodia Street, 8, Lviv, 79005, Ukraine}
\affiliation{$^2$College of Physics and International Center of Future Science of Jilin University, Qianjin Street 2699, Changchun, 130012, People's Republic of China}
\date{\today}

\abstract{We study the density distribution of the minimally-coupled scalar field dark energy inside a neutron star. The dark energy is considered in the hydrodynamical representation as a perfect fluid with three parameters (background density, equation of state, and effective sound speed). 
The neutron star matter is modeled with three unified equations of state, developed by the Brussels-Montreal group.
With the calculated density distribution of the dark energy inside a neutron star (and its dependence on the dark energy parameters) we investigate how its presence impacts the macroscopic characteristics and the value of the mass limit for neutron stars. From this impact we derive the possible constrains on the effective speed of sound of dark energy with the help of maximal known masses of observed neutron stars.
In this approach, we have found, that the squared effective speed of sound can not be smaller than $\sim 10^{-2}$ in units of squared speed of light.}

\maketitle
\section{Introduction}
The local behavior of the perturbed dark energy has become an object of study in a series of works of the last decade. The impact of perturbation and clusterization of dark energy on the matter dynamics was studied on clusters of galaxies and galaxy scales (for example, in \cite{Creminelli2010,Pace10,Tsizh2015,novosyadlyj_2016}) and on the astrophysical scale (\cite{Babichev2013,novosyadlyj_2014,smerechynskyi19,tsizh14,Kamiab11,Alavirad13,Pani11,Yazadjiev15,Sbisa20,Kpadonou15}). In general, studies have shown, that the local impact of minimally coupled perturbed dark energy is mostly negligible, except some specific value regions of its parameters space, which can be used for ruling out these values. 

In particular, the dark energy models as well as the modified gravity theories are tested as an altering factor of the compact objects' properties. Usually, in these works the authors investigate how the gravitational potential is altered due to hidden component or alternative gravitation theory and how it changes the characteristics of a compact object. The comparison of the theoretical predictions with the corresponding observational data can give some constraint on the value of parameters of the theory. There are also a number of papers where the probing objects are, in particular, neutron stars (NS). For example, the gravitational aether theory is tested with NSs in \cite{Kamiab11}. In \cite{Alavirad13} the effect of a logarithmic f(R) theory on relativistic stars is studied. Similarly, the alternative theories of gravity are tested by NSs in \cite{Pani11} (Einstein-Dilaton-Gauss-Bonnet gravity), \cite{Yazadjiev15,Sbisa20} (R-squared gravity) and \cite{Kpadonou15} (f(T) gravity).

The stationary accretion of dark energy onto the Schwarzschild black hole was studied in \cite{Babichev2013}. We studied the static solutions of dark energy dynamical equations in the vicinity of compact objects and found that only relativistic objects with the lowest ratio ``radius to gravitational radius'' disturb the density of dynamical dark energy noticeably \cite{novosyadlyj_2014}. In the paper \cite{smerechynskyi19} we have investigated how the dynamical dark energy inside white dwarfs can change their mass-radius relation and have found that the squared effective speed of sound $c_s^2$ of dark energy must be larger than $\sim 5 \cdot 10^{-4}$ in order to satisfy the observed mass-radius relation for white dwarfs. In the paper \cite{tsizh14}, we have used the uncertainty of determination of gravitating mass in the Solar system as
an upper limit on the amount of clustered dark energy
and obtained the similar constraints for the value of $c_s^2 \geq 2 \cdot 10^{-4}$. 
There are plenty of works which constrain the dark matter
parameters and theories of gravity based on the reduction of NS maximal mass caused by the accumulation of the dark matter in its interior (see \cite{Lavallaz10, Guver13, Ellis18,Ivanytskyi20}). In particular, authors of \cite{Deliyergiyev19} analyzed compact objects that contain dark matter admixed with ordinary matter made of neutron star and white dwarf materials and dependence of maximum radius of such objects on dark matter properties. In this work we use a similar approach towards another part of the dark sector -- minimally coupled dark energy.

The paper has the following structure. In Section~\ref{sect_2} we describe the equations of state for NS matter and hence present the model which governs the matter distribution inside NS. Section~\ref{sect_3} is devoted to the description of the dark energy model and the radial distribution of the dark energy inside NS and its dependence on the dark energy parameters. In Section~\ref{sect_4} we deduce the constraint on the effective speed of sound for the dark energy using the NSs and in Section~\ref{sect_5} we give our conclusions.

\section{Equation of state for the neutron star matter }
\label{sect_2}
The NSs exist due to the pressure of the degenerate gas of Fermi-particles, neutrons, much like the electrons in the case of white dwarfs, though, nuclear forces play an important role in the former ones. 
The state of the matter inside white dwarfs is well known, but we are still uncertain about the composition and equation of state (EoS) for the matter at the higher densities that correspond to the NS interiors \cite{haensel_book_2007}.

Based on a given EoS one can yield the maximum mass of NS configuration, the so-called Tolman-Oppenheimer-Volkoff mass limit \cite{Tolman, Oppenheimer}.
The last one is similar to the Chandrasekhar mass limit \cite{Chandrasekhar_1931a, Chandrasekhar_1931b, Chandrasekhar_1935}, but strongly depends on the incorporated physics, resulting in different stiffness of EoS.
The maximum mass limit for NS is approximately $1.5 M_{\odot}$ for soft equations of state and reaches $3 M_{\odot}$ in the case of stiff ones \cite{bombaci_1996, alsing_2018}.
The maximal known mass of observed NSs is a crucial value 
for testing
different EoSs
(see, for example, \cite{alsing_2018, zhou_2019,annala_2018,zhang_2018,zhang_2019}).

For the description of the NS interior we have exploited three unified EoSs developed by the Brussels-Montreal group \cite{goriely_2010, pearson_2011, pearson_2012,potekhin_2013}. 
Despite the matter being under different physical conditions and states, such an EoS is valid throughout all parts of a neutron star -- from the outer envelope to its crust and core.
We will use the same denotations BSk19, BSk20 and BSk21 for the considered EoSs as in \cite{potekhin_2013}.
They differ by its stiffness, BSk21 is the stiffest one and BSk19 is the softest one.

The analytical representation of the equations of state with variables $\xi=\lg(\rho_m/\textrm{g}\cdot\textrm{cm}^{-3})$ and
$\zeta = \lg(p_m/\textrm{dyn}\cdot\textrm{cm}^{-2})$ ($\lg$ denotes $\log_{10}$) is defined by the following parametrization:
\begin{eqnarray}
\begin{aligned}
  \zeta =
    \frac{a_1+a_2\xi+a_3\xi^3}{1+a_4\,\xi}\,
        \frac{1}{\exp\left[a_5(\xi-a_6)\right]+1}
     +\frac{a_7+a_8\xi}{\exp\left[a_9(a_6-\xi)\right]+1}
     + \frac{a_{10}+a_{11}\xi}{\exp\left[a_{12}(a_{13}-\xi)\right]+1}
\\
     + \frac{a_{14}+a_{15}\xi}{\exp\left[a_{16}(a_{17}-\xi)\right]+1}
     + \frac{a_{18}}{1+ [a_{19}\,(\xi-a_{20})]^2}
     + \frac{a_{21}}{1+ [a_{22}\,(\xi-a_{23})]^2}
     ,
\end{aligned}
\label{fit.P}
\end{eqnarray}
where $p_m$ and $\rho_m$ are the local pressure and density of the NS (or baryonic) matter, respectively.
It is the same for three considered cases, only the values of coefficients $a_1$-$a_{23}$ are different (see \cite{potekhin_2013}).
Such representation of EoS simplifies its usage for the consideration of the inner structure of NSs. The fitting procedure introduces the errors of the macroscopic NS characteristics, but they are far below the observational uncertainties \cite{potekhin_2013}.

\begin{figure}

\centering\includegraphics[width=.6\textwidth]{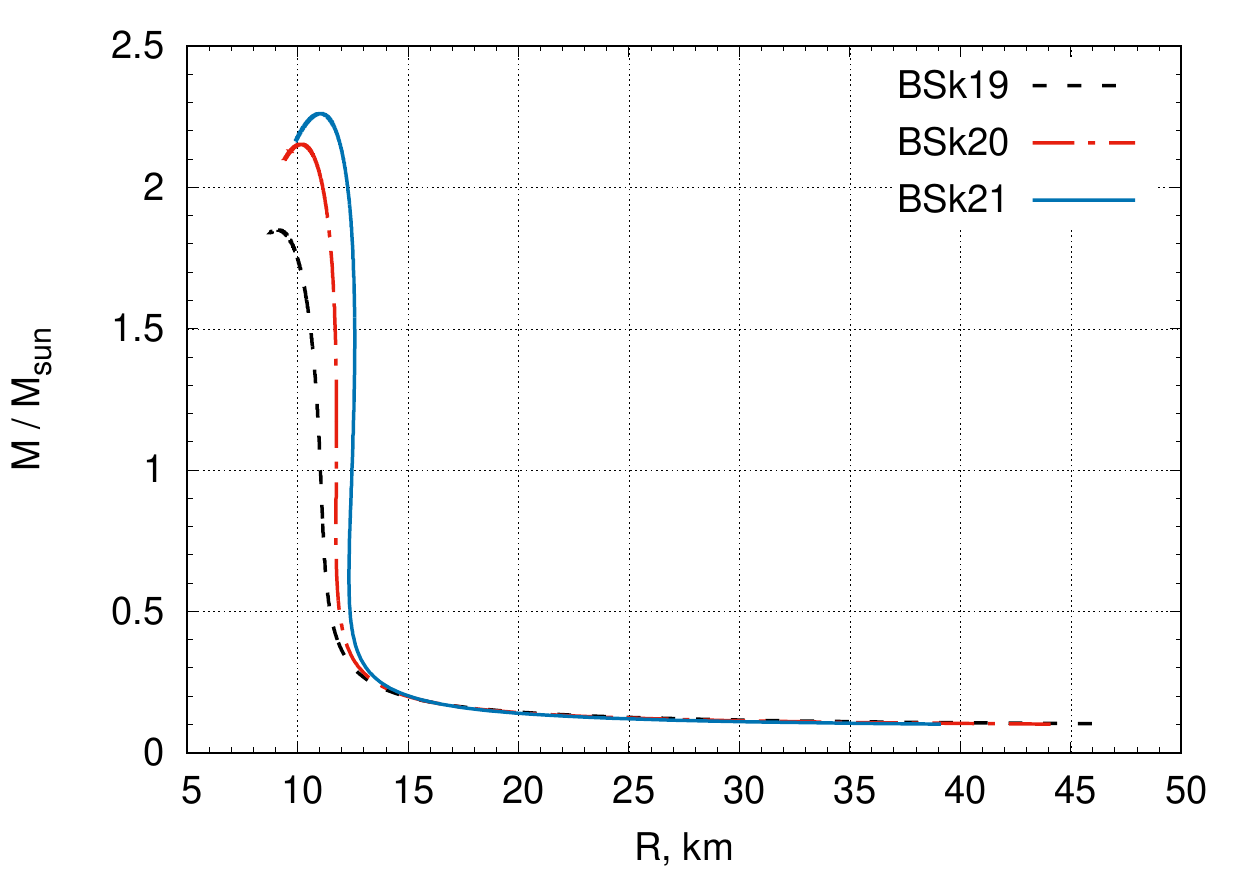}
\caption{Mass-radius relation for neutron stars with three considered equations of state.}
\label{m-r}
\end{figure}

Using these equations of state, one can solve the hydrostatic equilibrium equation to obtain the corresponding mass-radius relations, shown in Fig.~\ref{m-r}. 
As we can see, the stiffer the EoS is, the higher maximal mass of NS corresponds to it.
These values (in solar mass units) as well as the corresponding central densities $\rho_m(0)\equiv\rho_c$ of NSs are given in Table~\ref{m-r_tab}.

\begin{table}[h!]
\centering
\caption{The maximum masses and corresponding central densities of neutron stars for three equation of states BSk19, BSk20, BSk21 \cite{potekhin_2013}.}
\begin{tabular}{ccc}
\hline
\hline
EoS&$M_{max}/M_{\odot}$& $\rho_c$, $10^{15}$ g/cm$^3$\\
\hline
BSk19& $1.86$& $3.48$ \\
\hline
BSk20& $2.16$& $2.69$ \\
\hline
BSk21& $2.27$& $2.27$ \\
\hline
\hline
\end{tabular}
\label{m-r_tab}
\end{table}

The maximal known masses for observed NSs are: $2.14^{+0.10}_{-0.09}  M_{\odot}$ for PSR J0740+6620 (here and below the confidence interval is 68.3\%), obtained from the measurement of the relativistic Shapiro delay \cite{cromartie_2019}; and $2.27_{-0.15}^{+0.17} M_{\odot}$ for PSR J2215+5135, yielded by the simultaneous fitting of radial velocity curves and three-band light curves \cite{linares_2018, linares_2019}.
It is worth mentioning that there are objects which are reported to have masses even larger. For instance, pulsar PSR B1957+20 with mass $\sim 2.4 M_{\odot}$ \cite{kerkwijk2010}, but systematic uncertainties of the mass determination are large. It is argued, that the accuracy of this measurement is partly limited by optical flares and variable emission lines of companion's stellar wind  \cite{romani_2015,linares_2019}. There are evidences indicating significantly lower mass of that pulsar ($\sim1.8 M_{\odot}$ \cite{horvath_2017,luo_2020}).
Thus, to date, the maximum mass of observed NSs, in general, agrees with estimates on the upper bound of NS masses made within different approaches \cite{alsing_2018, shibata_2019,lawrence_2015,fryer_2015,margalit_2017,rezzolla_2018,ruiz_2018,abbott_2017}.

One can infer from the comparison of theoretical maximum masses from Table~\ref{m-r_tab} and observed ones, that the maximum mass values predicted by the equations of state BSk20 and BSk21 are inside the range inferred from observations and BSk19 does not allow the existence of such massive NSs. Probably, there is a discrepancy for BSk20 EoS and PSR J2215+5135 mass, if no evidence of significant uncertainty of its mass determination is found. In the following sections we will use all three mentioned equations of state, bearing in mind that BSk19 can not explain the existence of the most massive among known NSs and we consider it only for the sake of consistency and analysis.

\section{Dark energy inside a neutron star}
\label{sect_3}
\subsection{Dark energy model}
It was shown that the minimally-coupled scalar field model of dark energy with barotropic EoS
\begin{equation}
\label{eq1a}
p_{de}=w(\rho_{de})c^2\rho_{de},
\end{equation}
where $p_{de}$ and $\rho_{de}$ are pressure and density of dark energy, respectively, can agglomerate inside and in vicinity of a compact object when the EoS parameter $w$ and the squared effective speed of sound $c_s^2$ ($c_s$ in the units of speed of light) are related as \cite{Babichev2013, novosyadlyj_2014} 
\begin{equation}
w=c_s^2-(c_s^2-w_{\infty})\frac{\rho_{\infty}}{\rho_{de}}.\label{w-rho}
\end{equation}
Here $\rho_{\infty}$ is the background density of dark energy (at $r\rightarrow \infty$), which in our case is equal to $10^{-23}$g/cm$^3$ \cite{novosyadlyj_2013}. The value of background dark energy density was chosen to be higher than the cosmological value in $\Lambda$CDM model. The reason for this is that dark energy is assumed to undergo clusterization process along with dark matter during initial perturbation growth. We considered only the quintessence type of dark energy with $w_{\infty}>-1$. The model, described by equation (\ref{w-rho}), implies the constant effective speed of sound of dark energy.

The hydrodynamical representation of the scalar field dark energy as a perfect or imperfect fluid with barotropic EoS is usually used in cosmology. The Lagrangian of the field $\mathcal{L}(X,U)$ with kinetic term $X$ and potential $U$, is connected to phenomenological hydrodynamical quantities as follows \cite{armendariz-picon_1999}
$$ c^2 \rho_{de} = 2X\mathcal{L}_{,X} - \mathcal{L},\quad p_{de} =  \mathcal{L}, \quad
w = \frac{p_{de}}{c^2\rho_{de}} = \frac{\mathcal{L}}{2X\mathcal{L}_{,X}}, \quad c_s^2 = \frac{\delta p_{de}}{c^2\delta\rho_{de}} = \frac{\mathcal{L}_{,X}}{2X\mathcal{L}_{,XX} - \mathcal{L}_{,X}}. $$
The scalar field dark energy with conditions $c_s^2=const>0$ and $w<0$ in stationary Minkowski or Schwarzschild world is governed by the Klein-Gordon or hydrodynamical continuity equations. In \cite{novosyadlyj_2014} we have shown how the eq. (3) is deduced in the framework of these conditions and in \cite{Sergijenko14} how the scalar field variables are related with hydrodynamical ones for this dark energy model.

\subsection{Distribution of the dark energy inside a neutron star}

In order to estimate the influence of the dynamical dark energy on the equilibrium condition in a NS we suppose that it is the non-rotational non-magnetic star which is in static equilibrium: the pressure gradient of baryon matter balances the gravitational attraction of the total mass in a given sphere in the star, as well as the pressure gradient of the dark energy balances the gravitational attraction of the same mass. The Einstein and conservation law equations for minimally coupled baryonic and dark energy are used.
Therefore, we considered a spherically symmetric object for which the space-time metric can be written in the form
\begin{equation}
\label{eq2}
ds^2=e^{\nu(r)}c^2d\tau^2-e^{\lambda(r)}dr^2-r^2\left(d\theta^2+\sin^2{\theta}d\varphi^2\right).
\end{equation}

If we limit ourselves to the case of static configuration of dark energy inside a NS, the components of metric will not depend on time and can be obtained from the Einstein equations with the boundary condition $\lambda(r=0) = 0$ \cite{novosyadlyj_2014}
\begin{equation}
\label{metric_func}
\begin{aligned}
e^{-\lambda(r)}&=1-\frac{8\pi G}{c^2r}\int\limits_0^r \left[\rho_m(r')+\rho_{de}(r')\right]r'^2dr',\\
\nu(r)+\lambda(r)&=-\frac{8\pi G}{c^2}\int\limits_r^{\infty} \left[\rho_m(r')+\rho_{de}(r')+\frac{p_m(r')+p_{de}(r')}{c^2}\right] e^{\lambda(r')}r'dr'.
\end{aligned}
\end{equation}
Here $\rho_m$, $p_m$ are the local density and pressure of baryonic matter and $\rho_{de}$, $p_{de}$ denote the corresponding characteristics of dark energy. Other boundary conditions are the following: $\nu(\infty) = -\lambda(\infty)$, $\rho_m(R_+)=0$, $\rho_{de}(\infty)=\rho_{\infty}$.

With the metric functions given in (\ref{metric_func}) we numerically solved the equilibrium equations for both baryonic matter and dark energy
\begin{eqnarray}
\label{equilibr_eq}
\begin{aligned}
\frac{dp_m}{dr}+\frac12(\rho_m c^2+p_m)\frac{d\nu}{dr}=0,\\
\frac{dp_{de}}{dr}+\frac12(\rho_{de} c^2+p_{de})\frac{d\nu}{dr}=0,
\end{aligned}
\end{eqnarray}
applying the iterative procedure. On the initial step we evaluated the gravitational potential without the dark energy influence, and, thereafter, found the distribution of the dark energy in such potential. Then we solved the system of equations for baryonic matter and dark energy densities and their joint gravitational potential, and compared the results with ones obtained in the previous step. Then we re-evaluated the distribution of dark energy and baryonic matter in the new potential. Such procedure was repeated until the solutions converged or the iteration limit exceeded (for more details see~\cite{smerechynskyi19}). 

\begin{figure}[h!]
\begin{minipage}{.32\textwidth}
\includegraphics[width=1\textwidth]{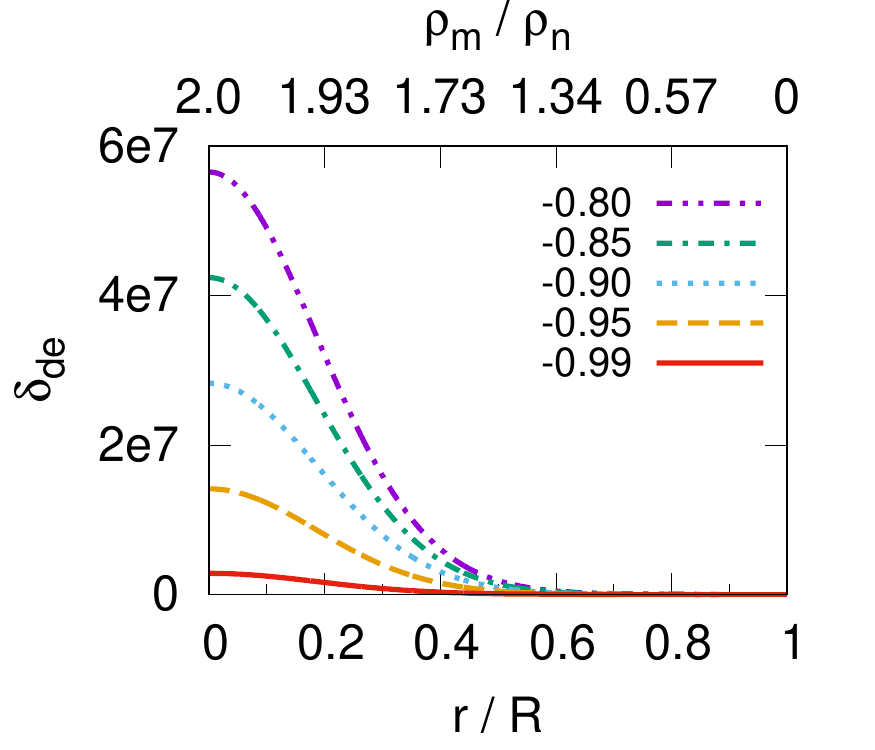}
\centering{(a)}
\end{minipage}
\begin{minipage}{.32\textwidth}
\includegraphics[width=1\textwidth]{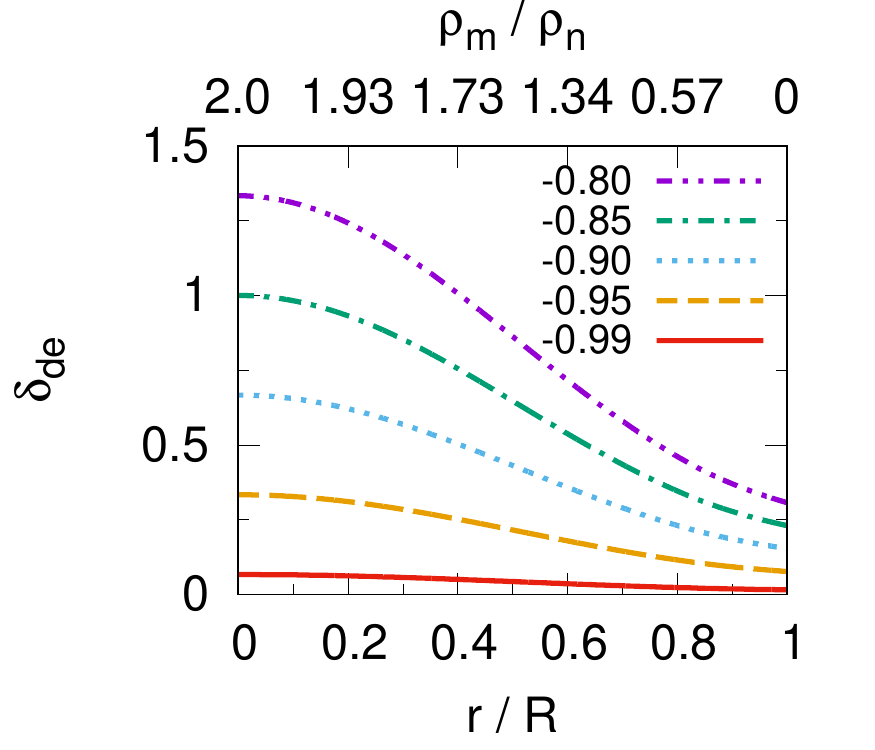}
\centering{(b)}
\end{minipage}
\begin{minipage}{.32\textwidth}
\includegraphics[width=1\textwidth]{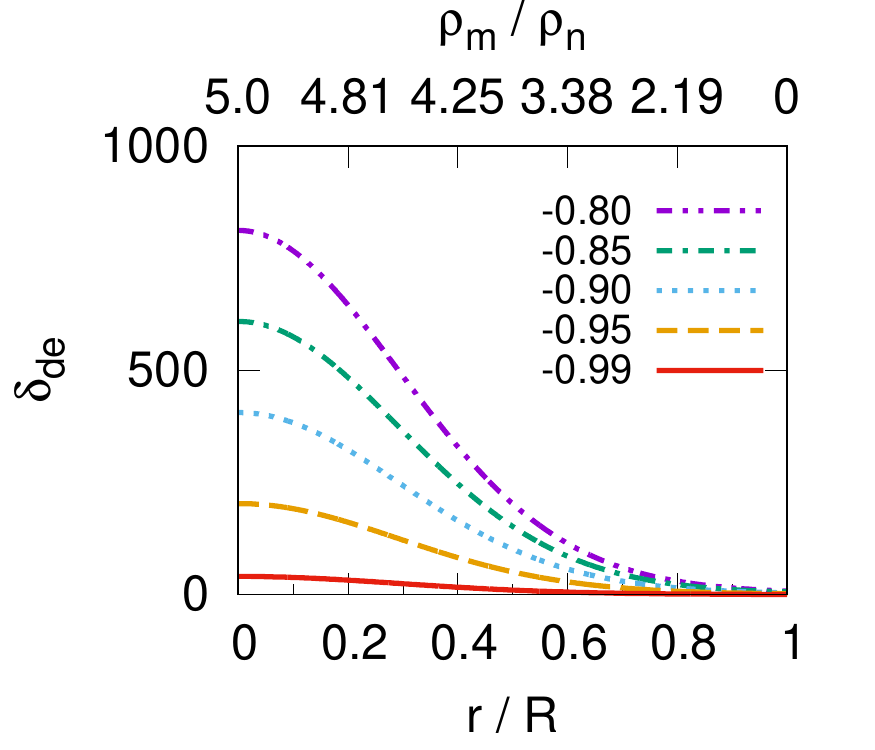}
\centering{(c)}
\end{minipage}
\caption{The relative deviation $\delta_{de} (r) = (\rho_{de}(r)-\rho_{\infty})/\rho_{\infty}$ of dark energy density as a function of radial coordinate $r$ inside a neutron star with radius $R$ for the case of BSk20 EoS: (a) $\rho_c = 2\rho_n$, $c_s^2=0.01$; (b) $\rho_c = 2\rho_n$, $c_s^2=0.1$; (c) $\rho_c = 5\rho_n$, $c_s^2=0.1$. Line types correspond to different values of the parameter $w_{\infty}$: from $-0.80$ for top curve to $-0.99$ for bottom curve. The upper x-axis corresponds to matter density at given radial coordinates.}
\label{rho_r_20}
\end{figure}

The solutions of the system of equations (\ref{metric_func}--\ref{equilibr_eq}) for the dark energy component are shown in Fig.~\ref{rho_r_20} in the form of radial dependence of the relative deviation of the dark energy density $\delta_{de}(r)$ inside a star. The results correspond to BSk20 EoS for two different values of effective speed of sound $c_s^2 = 0.01$ and $0.1$, and central density of the NS matter $\rho_c=2\rho_n$ and $5\rho_n$, where $\rho_n=2.8\cdot 10^{14}$~g/cm$^3$ is the so called normal nuclear density \cite{haensel_book_2007}. 
The line types correspond to the different values of the parameter $w_{\infty}$ given in the figure.
The upper x-axis corresponds to the matter density at the given radial coordinates.

One can see, that the relative deviation of the dark energy density in NS is very sensitive to the value of $c_s^2$ and is increasing as the latter one is decreasing. It follows from the comparison of Figs.~\ref{rho_r_20}a and \ref{rho_r_20}b, corresponding to the same value of $\rho_c$. $\delta_{de}(r)$ increases also with increasing central density of baryonic matter $\rho_c$ at constant effective speed of sound of dark energy $c_s^2$ (Figs.~\ref{rho_r_20}b and \ref{rho_r_20}c). 

Also, one can infer that lowering $w_{\infty}$ causes a smaller deviation of the dark energy density from the background one. This makes no surprise: it is well known that the dark energy with $w_{\infty}=-1$ is not perturbed at all, so one would expect the deviation to be smaller as $w_{\infty}$ approaches -1 and vanishes at $w_{\infty}=-1$.

\begin{figure}[h!]
\begin{minipage}{.32\textwidth}
\includegraphics[width=1\textwidth]{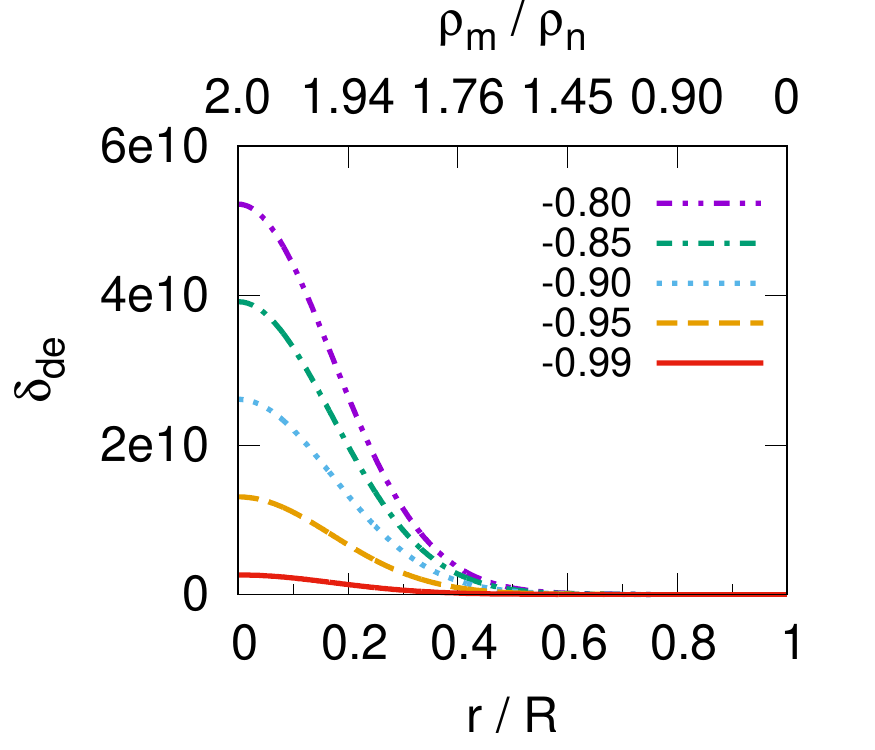}
\centering{(a)}
\end{minipage}
\begin{minipage}{.32\textwidth}
\includegraphics[width=1\textwidth]{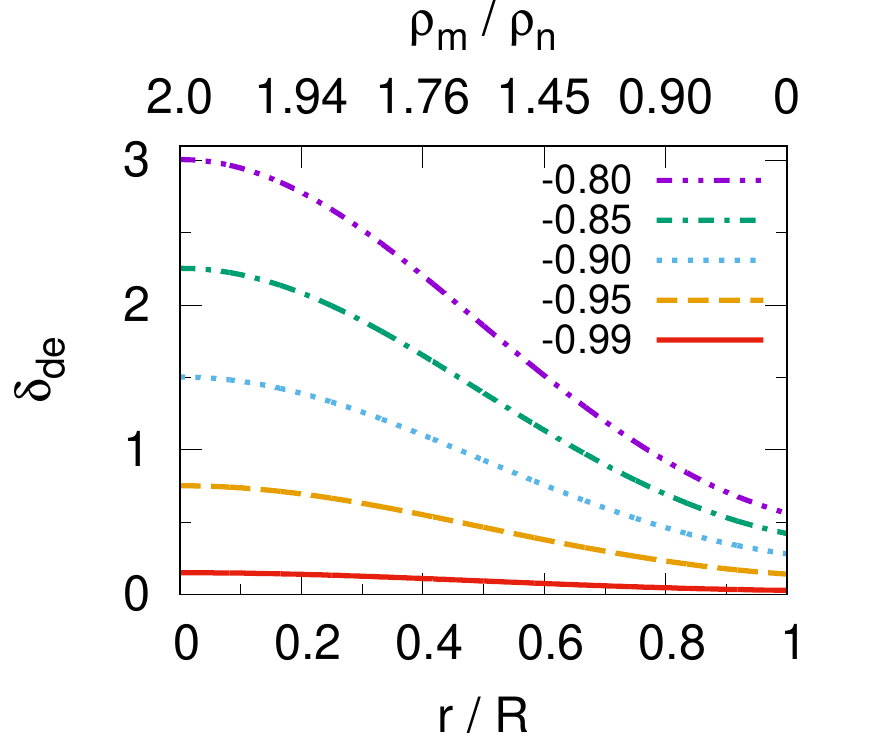}
\centering{(b)}
\end{minipage}
\begin{minipage}{.32\textwidth}
\includegraphics[width=1\textwidth]{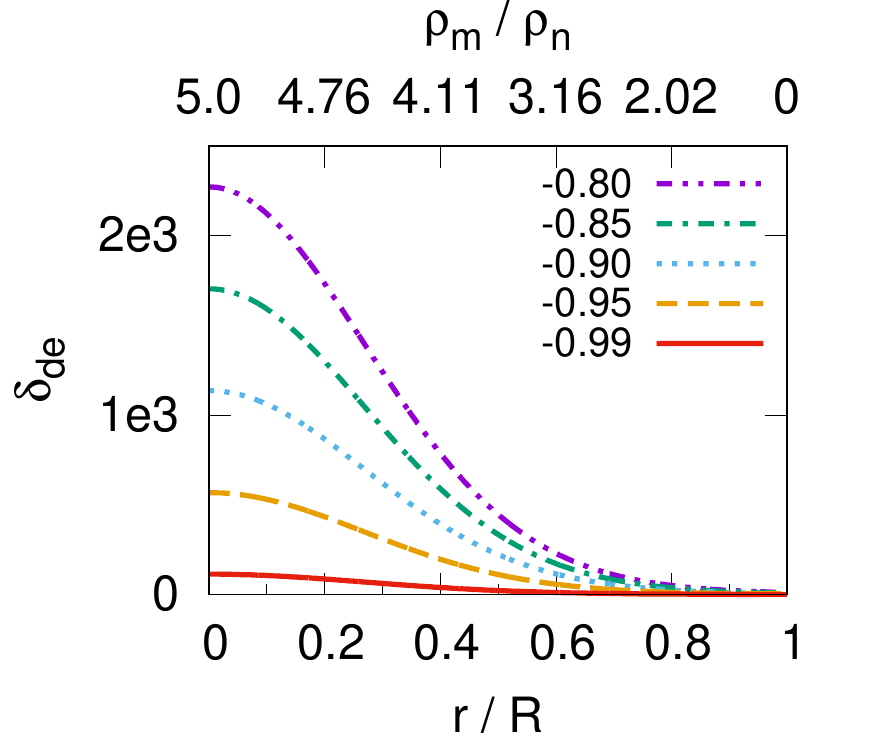}
\centering{(c)}
\end{minipage}
\caption{The same as in Fig.~\ref{rho_r_20}, but for the case of BSk21 equation of state.}
\label{rho_r_21}
\end{figure}

The solutions for the case of the stiffest of considered EoSs, namely BSk21, are illustrated in Fig.~\ref{rho_r_21}. Similar dependencies of the relative deviation of the dark energy density on parameters $w_{\infty}$, $c_s^2$ and $\rho_c$ can be inferred from Figs.~\ref{rho_r_21}a--\ref{rho_r_21}c. However, it should be noted that the amount of dark energy inside a NS is larger for the case of stiffer EoS assuming the same values of other parameters.

\section{Influence of dark energy on neutron star and parameter constraints}
\label{sect_4}

\begin{figure}[h!]
\centering\includegraphics[width=.6\textwidth]{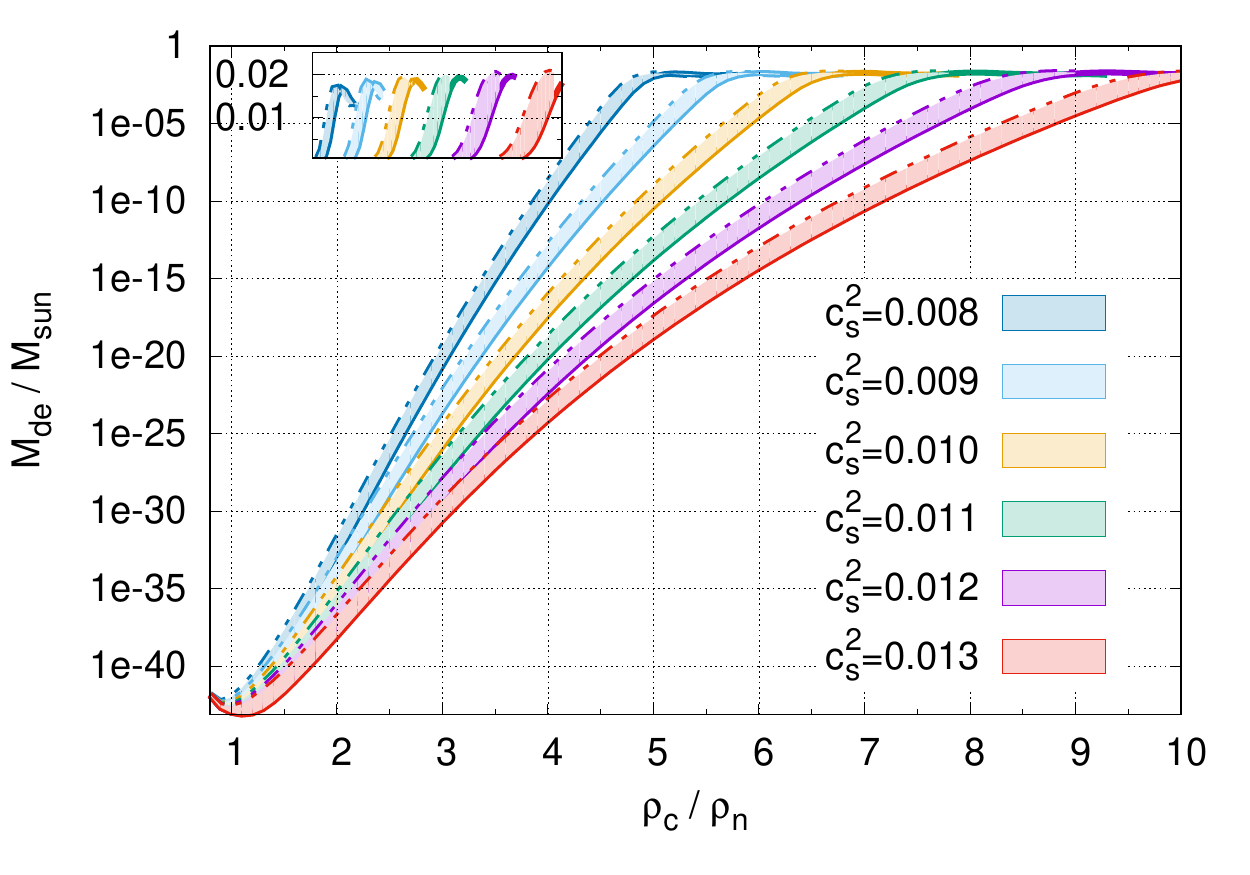}
\caption{The dark energy mass (in solar mass units and logarithmic scale) as a function of the neutron star central density for the case of BSk21 equation of state. The colors correspond to different values of $c_s^2$: from $0.008$ (top) to $0.013$ (bottom). The results for $w_{\infty}=-0.99$ are depicted with solid lines and ones for $w_{\infty}=-0.80$ are depicted with dash-dotted lines, shadowed area between these lines corresponds to the results for in-between values of $w_{\infty}$. Zoomed-in region of $M_{de}/M_{sun}\sim 10^{-2}$ is shown in the upper left corner of the figure in the linear scale.}
\label{m_de-rho}
\end{figure}

The density distributions of dark energy obtained in the previous section give us the possibility to calculate its total Lagrangian mass inside a star. Fig.~\ref{m_de-rho} illustrates the dark energy mass $M_{de}$ (in solar mass units) as a function of central density $\rho_c$ of baryonic matter for the values of $c_s^2$ ranging from $0.008$ to $0.013$ (depicted with different colors). The dependence on the parameter $w_{\infty}$ is shown by shadowed regions between solid lines corresponding to $w_{\infty}=-0.99$ and dash-dotted lines ($w_{\infty}=-0.80$). The dark energy mass rises steeply with central matter density with an exception of the region near $\rho_c/\rho_n\approx1$ and then reaches saturation level which is in the range of one percent of total mass for all considered values of $c_s^2$ and $w_{\infty}$ (see zoomed-in part of the figure in the upper left corner). The amount of dark energy is higher for smaller values of $c_s^2$ and larger values of $w_{\infty}$. Moreover, the results for $M_{de}$ are more sensitive to the change of $c_s^2$.

Aiming to constrain the parameters of dark energy we have studied its influence on the NS mass. The dark energy does not reveal itself until a certain value of central matter density is reached because of very strong dependence on the ratio of gravitational radius to stellar surface one\footnote{The ratio $M_{de}/M\ll1$ for objects with $R\ll r_g$ for $c_s^2>10^{-4}$ \cite{novosyadlyj_2014}, where $R$ is radius of object, $r_g$ is its gravitational radius. So, the dynamical dark energy practically does not influence the gravitational field in the normal stars.}. This value depends on the chosen EoS for the NS matter and the parameters of dark energy. Thus, in order to analyze the impact of each parameter, we have consecutively fixed all of them except one. As was mentioned above, in our calculations we adopted $\rho_{\infty} = 10^{-23}$g/cm$^3$, and this parameter was not changed at all.

\begin{figure}[h!]
\begin{minipage}{.51\textwidth}
\centering\includegraphics[width=\textwidth]{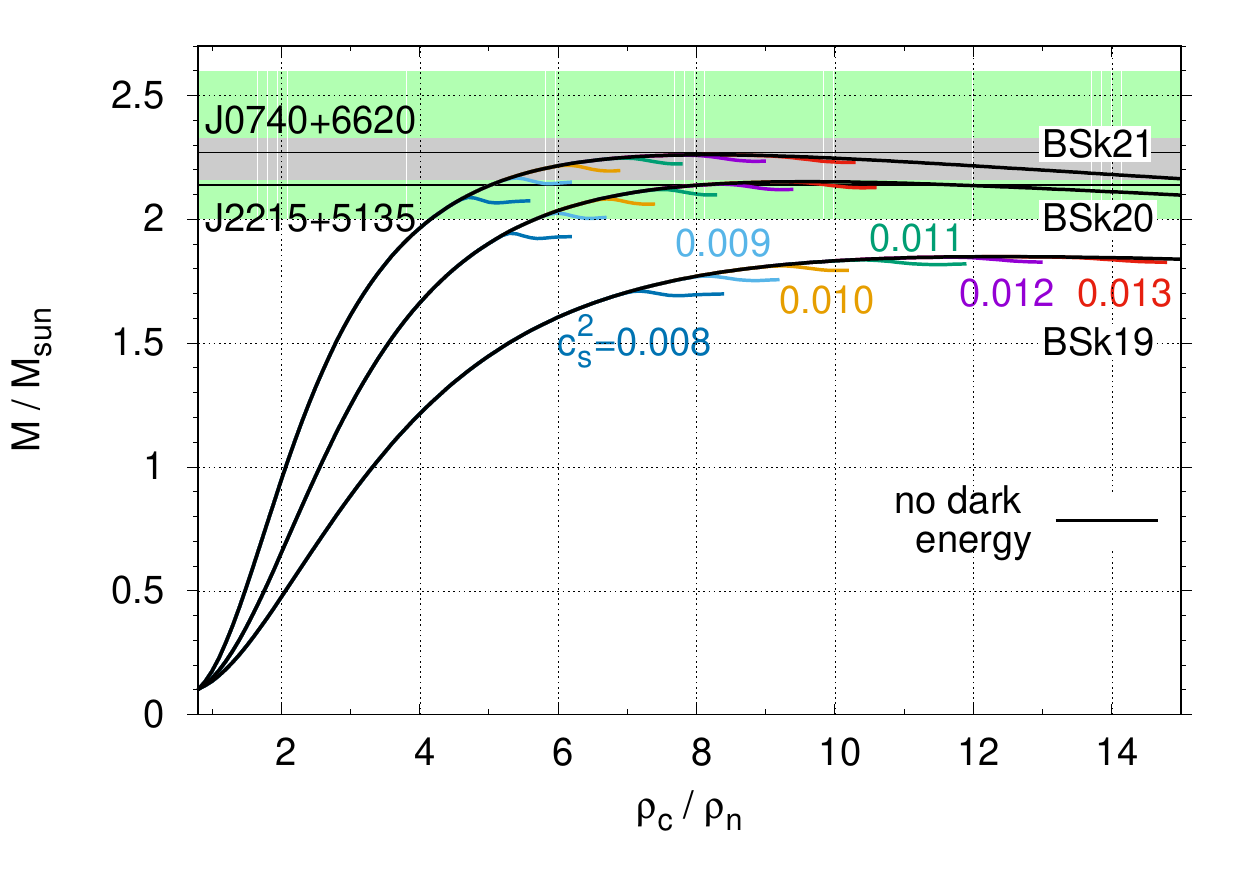}
\centering{(a)}
\end{minipage}
\begin{minipage}{.48\textwidth}
\centering\includegraphics[width=\textwidth]{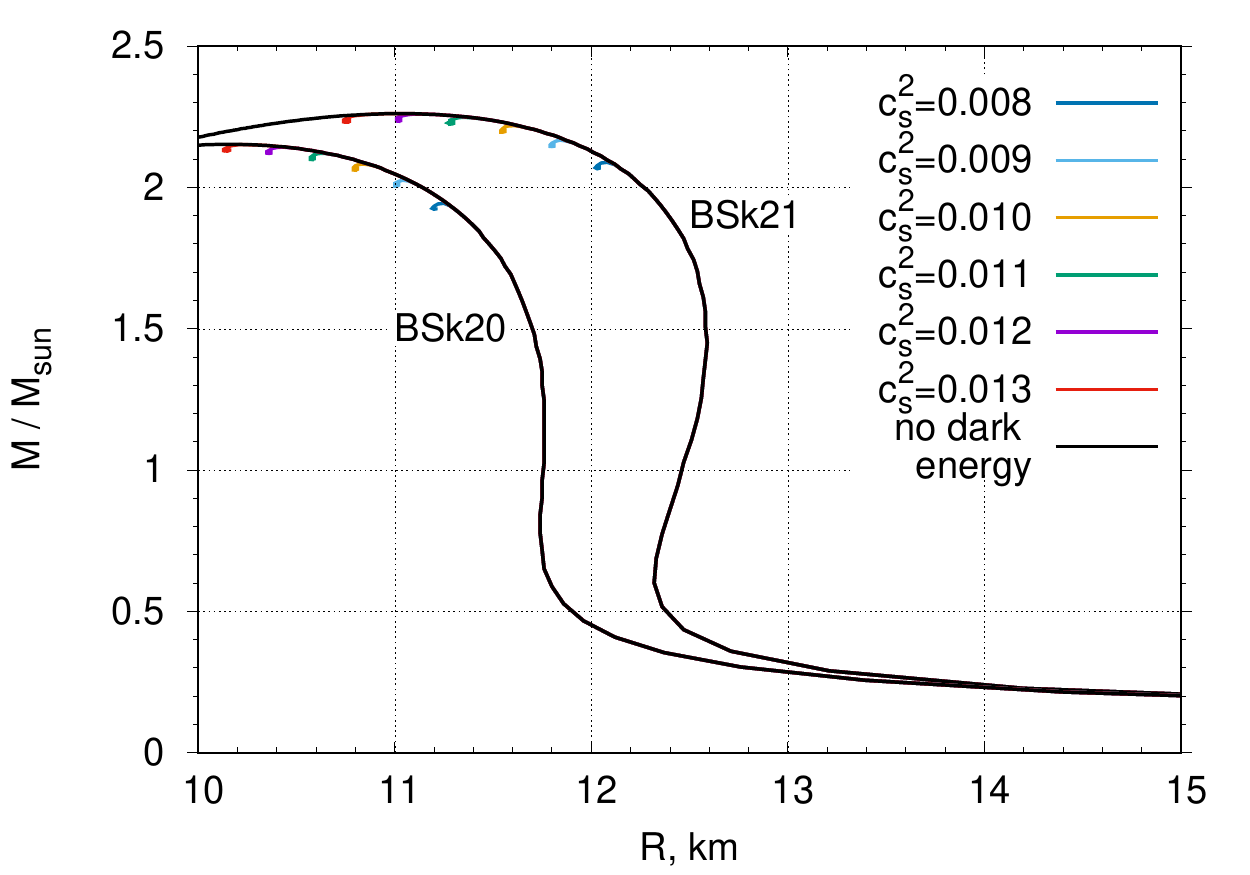}
\centering{(b)}
\end{minipage}
\caption{(a) The total mass of neutron star configuration (including the dark energy inside) as a function of its central matter density for three considered equations of state for NS matter and different values of $c_s^2$ and $w_{\infty}=-0.8$ for dark energy (see text for details); (b) Mass-radius relation for two equations of state of NS matter in models with dark energy with different values of $c_s^2$ (depicted with different colors) and without it (black solid lines).}
\label{m-rho}
\end{figure}

The total mass of a NS configuration (including the dark energy inside) as a function of its central matter density is shown in Fig.~\ref{m-rho}a for three considered equations of state (labeled respectively) and the same values of squared effective speed of sound $c_s^2$ as in Fig.~\ref{m_de-rho}, here the EoS parameter remained fixed ($w_{\infty}=-0.8$). Similarly to white dwarfs, the quintessence type of dark energy reduces the NS mass, acting matter-like, and contributing to the joint gravitational potential.
The corresponding mass-radius relations for two considered EoSs for NS matter are shown in Fig.~\ref{m-rho}b.

\begin{figure}[h!]
\begin{minipage}{.49\textwidth}
\centering\includegraphics[width=\textwidth]{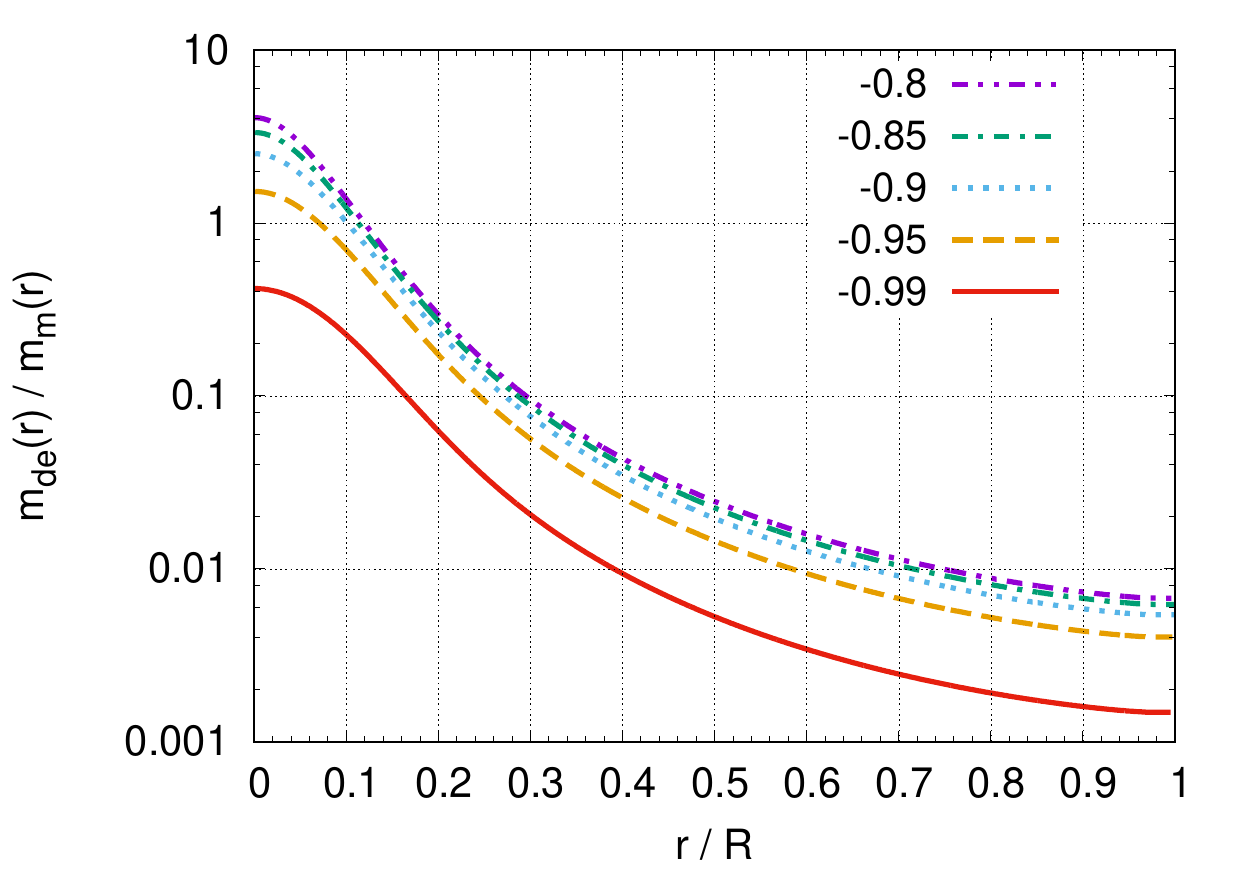}
\centering{(a)}
\end{minipage}
\begin{minipage}{.49\textwidth}
\centering\includegraphics[width=\textwidth]{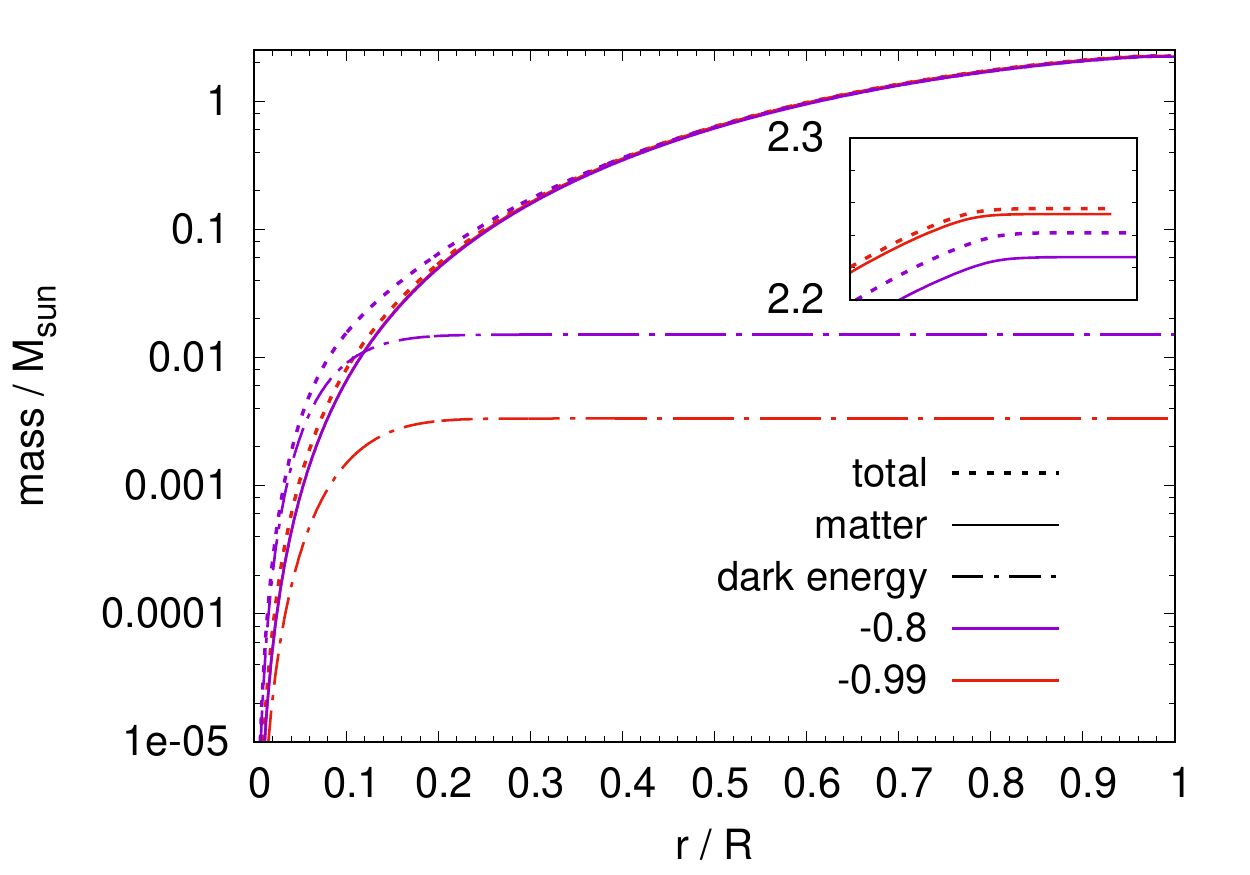}
\centering{(b)}
\end{minipage}
\caption{The dark energy to baryonic mass ratio $m_{de}(r)/m_m(r)$ as functions of dimensionless radial coordinate for given values of $w_{\infty}$ (depicted with different colors and line types); 
(b) The radial dependencies of dark energy mass (dash-dotted lines), mass of baryonic matter (solid lines) and total mass, which is the sum of previous two (dotted lines), in units of solar mass for two values of $w_{\infty}$ (depicted with different colors). Zoomed-in surface region $(0.95 - 1)$ $r/R$ is shown in the upper right corner. The results correspond to the choice of BSk21 EoS. Other parameters have values $\rho_c = 8.5\rho_n$, $c_s^2=0.012$.}
\label{m_de-m_m_r}
\end{figure}

As we can see from both figures, the influence of dark energy becomes crucial at some value of $\rho_c$ which depends on $c_s^2$.
At the central matter densities higher than this ``turn-off'' point, the dark energy content causes abrupt deviation from the model without the latter one (black solid lines). 
The amount of dark energy accumulated inside a neutron star with central matter density near the turn-off point can be seen in Fig.~\ref{m_de-m_m_r} for the model with $\rho_c = 8.5\rho_n$, $c_s^2=0.012$ and BSk21 EoS as an example. On the left panel (Fig.~\ref{m_de-m_m_r}a) the ratio between dark energy mass $m_{de}(r)$ and mass of baryonic matter $m_m(r)$ is shown as functions of dimensionless radial coordinate for given values of the parameter $w_{\infty}$ (depicted with different line styles and colors). Because of the concentration of dark energy towards the center (see Fig.~\ref{rho_r_21}), it dominates in the central part of the object (except the model with $w_{\infty}=-0.99$ for which $m_{de}<m_m$ even in stellar center). But with growing radial coordinate $r$, the ratio decreases and reaches the values less than 0.01 on the surface for all considered $w_{\infty}$. On the right panel (Fig.~\ref{m_de-m_m_r}b) we can see the radial dependencies of dark energy mass (dash-dotted lines), mass of NS matter (solid lines) and total mass, which is the sum of the previous two (dotted lines), in units of solar mass for two values of $w_{\infty}$ (depicted with different colors). 
The baryonic masses are only less than 1 percent lower at $r=R$ than total ones (see zoomed-in region in the upper right corner of the figure) and both baryonic and total masses are slightly lower for the model with $w_{\infty}=-0.8$, while the amount of dark energy is higher in this case. Therefore, more than 99 percent of the total mass of stable NS consists of baryonic mass.

Returning to Fig.~\ref{m-rho}, at the higher central matter densities the amount of dark energy is so high, that the mass of the matter drops and in this region of $\rho_c$ there are no stable equilibrium configurations. The pressure of baryonic matter can no more resist gravitation force from potential strengthened by dark energy. Therefore, for a given value of $c_s^2$ we obtain a corresponding existence region for central matter densities of NSs, roughly constrained by this turn-off point.

This fact can be used to find a lower bound on the parameter $c_s^2$, but for that purpose one needs the observational data on NS masses. With grey color in Fig.~\ref{m-rho}a we depicted the upper bound range for maximum NS mass given in papers~\cite{margalit_2017,rezzolla_2018,ruiz_2018}, the results of which are based on the observations of the binary NS merger GW170817 \cite{abbott_2017}. The masses of the most massive NSs, PSR J0740+6620 and PSR J2215+5135, are indicated with labeled straight lines. In addition, 90\% credible region for maximum mass, obtained in paper~\cite{alsing_2018} with help of Bayesian model selection analysis for NS mass distribution, is shown as the green filled area.

One can immediately conclude from Fig.~\ref{m-rho}a, that the accuracy of NS mass determination or/and the increase of the number of NSs with known masses are crucial for setting a tight constraint on the effective sound speed of dark energy.  Considering the upper bounds for maximum NS mass (grey region) we found that $c_s^2$ should be larger than $0.009$ in the case of BSk21 EoS, and $0.013$ for BSk20.
Assuming the mass estimation of PSR J0740+6620 is reliable, we can constrain $c_s^2\gtrsim0.012$ for BSk21. It should be mentioned, that these constraints depend on the considered EoS and the accuracy of the mass determination. In general, for a fixed NS model, the higher the maximum mass of NSs is found, the higher is the lower limit for $c_s^2$ value. On the other hand, such high maximum masses make possible and at some point even require stiffer equations of state (to allow their existence), and in that case the opposite is true: the model with stiffer EoS of NS lowers the bound on $c_s^2$. 

\begin{figure}[h!]
\begin{minipage}[t]{.49\textwidth}
\centering\includegraphics[width=1\textwidth]{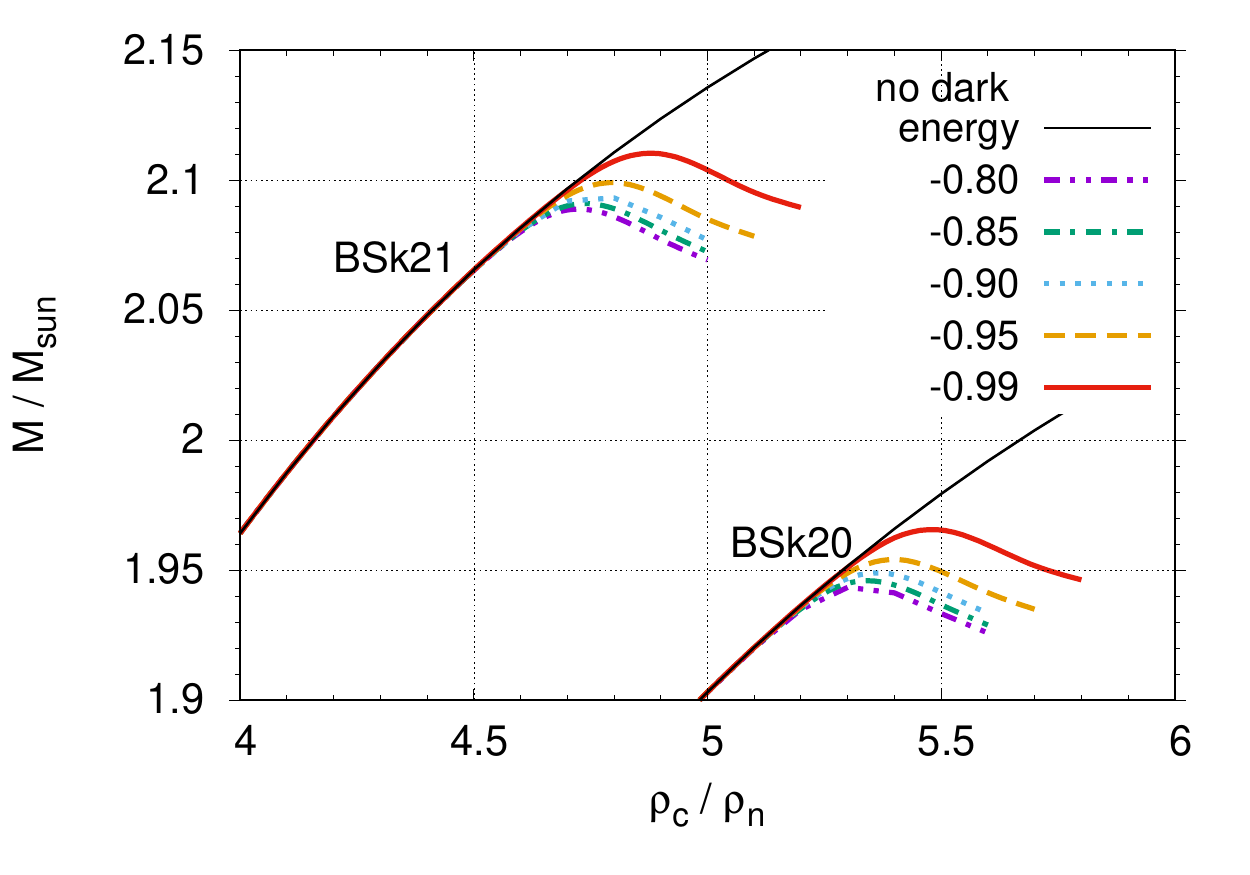}
\centering{(a)}
\end{minipage}
\begin{minipage}[t]{.49\textwidth}
\centering\includegraphics[width=1\textwidth]{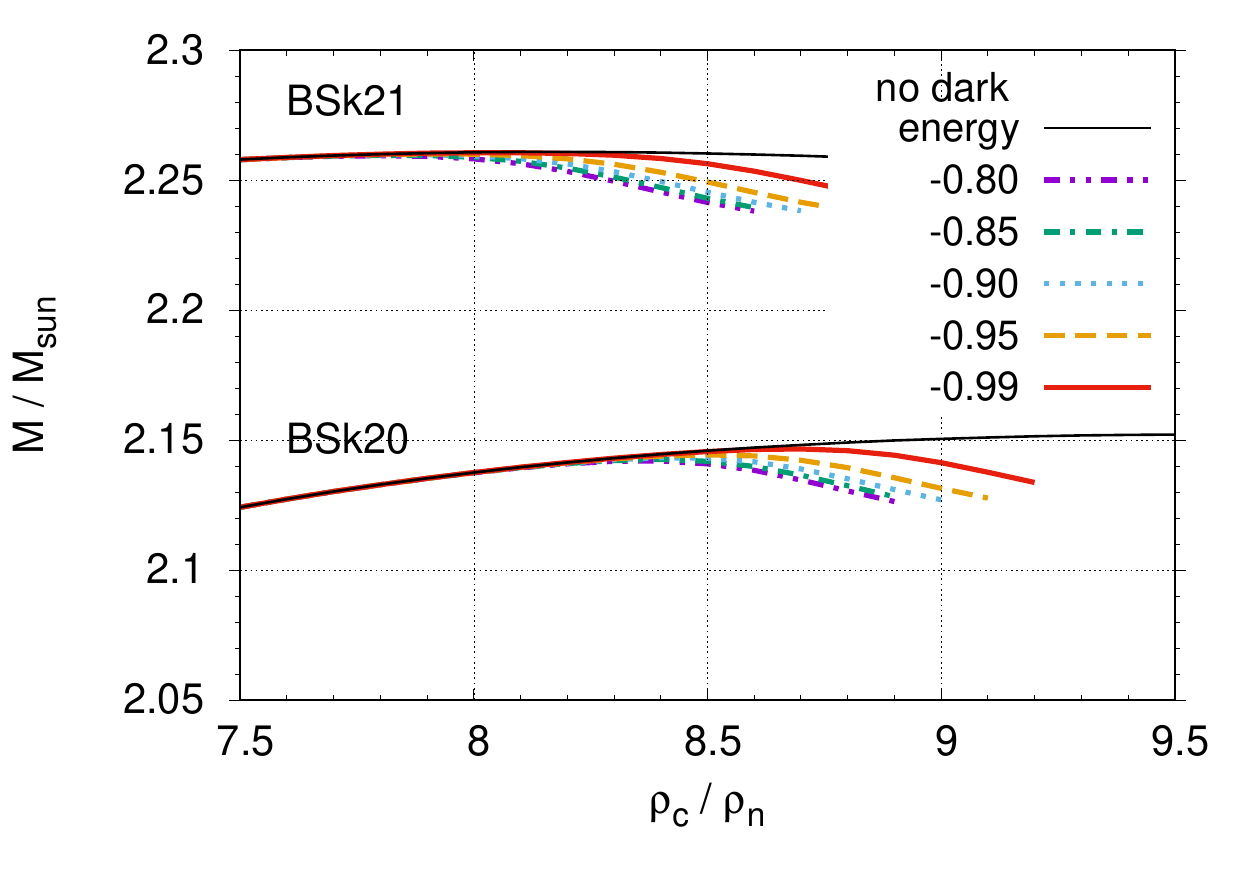}
\centering{(b)}
\end{minipage}
\caption{Similar as in Figure~\ref{m-rho}a, but for different values of $w_{\infty}$ and fixed $c_s^2$ (see text for details):
 (a) $c_s^2=0.009$; (b) $c_s^2=0.012$.}
\label{m-rho-cs2}
\end{figure}

As a next step, we investigated how the location of a turn-off point depends on parameter $w_{\infty}$ at fixed value of $c_s^2$. The results for the total mass of NS as a function of $\rho_c$ for $c_s^2=0.009$ are illustrated in Fig.~\ref{m-rho-cs2}a and for $c_s^2=0.012$ -- in Fig.~\ref{m-rho-cs2}b. As in Fig.~\ref{m-rho}a, the results of the model without dark energy are shown with black solid lines. We considered the values of $w_{\infty}$ ranging from $-0.80$ to $-0.99$. For these values of $w_{\infty}$, the central matter density $\rho_c$ corresponding to the turn-off point varies approximately on 3.9\% and 4.0\% for BSk21 and BSk20, respectively, in the case $c_s^2=0.009$, while 3.5\% and 5.0\% are corresponding variations for the case $c_s^2=0.012$. The values 
of the turn-off point for different values of dark energy parameters $c_s^2$ and $w_{\infty}$ and three considered EoSs
are given in Table~\ref{rho_cs_w_tab}. Thus, one can infer from Fig.~\ref{m-rho-cs2} and Table~\ref{rho_cs_w_tab} that the results are 
less sensitive to the choice of $w_{\infty}$ than $c_s^2$
for all considered equations of state.

\begin{table}[h!]
\centering
\caption{The values of turn-off point -- the central matter density of NS $\rho_c$ (in units of normal nuclear density $\rho_n$) at which the dark energy with different values of $c_s^2$ and $w_{\infty}$ reveals itself on the mass -- central density dependence (see Figs.~\ref{m-rho}a and \ref{m-rho-cs2}). The results are given for three considered equations of state BSk19, BSk20 and BSk21.}
\begin{tabular}{c|cccccc}
\hline
\hline
\multicolumn{7}{c}{BSk21}\\
\hline
\backslashbox{$w_{\infty}$}{$c_s^2$} & $0.008$ & $0.009$ & $0.010$ & $0.011$ & $0.012$ & $0.013$\\
\hline
$-0.80$ & $4.5$ & $5.1$ & $5.8$ & $6.6$ & $7.6$ & $8.6$\\
$-0.90$& $4.5$ & $5.2$ & $5.9$ & $6.7$ & $7.6$ & $8.7$\\
$-0.99$& $4.6$ & $5.3$ & $6.0$ & $6.9$ & $7.9$ & $9.0$\\
\hline
\hline
\multicolumn{7}{c}{BSk20}\\
\hline
\backslashbox{$w_{\infty}$}{$c_s^2$} & $0.008$ & $0.009$ & $0.010$ & $0.011$ & $0.012$ & $0.013$\\
\hline
$-0.80$ & $5.1$ & $5.7$ & $6.4$ & $7.1$ & $7.9$ & $8.9$\\
$-0.90$& $5.1$ & $5.7$ & $6.4$ & $7.2$ & $8.0$ & $9.0$\\
$-0.99$& $5.2$ & $5.9$ & $6.6$ & $7.4$ & $8.3$ & $9.3$\\
\hline
\hline
\multicolumn{7}{c}{BSk19}\\
\hline
\backslashbox{$w_{\infty}$}{$c_s^2$} & $0.008$ & $0.009$ & $0.010$ & $0.011$ & $0.012$ & $0.013$\\
\hline
$-0.80$ & $6.8$ & $7.7$ & $8.7$ & $9.8$ & $11.1$ & $12.6$\\
$-0.90$& $6.8$ & $7.8$ & $8.8$ & $10.0$ & $11.3$ & $12.7$\\
$-0.99$& $7.0$ & $8.0$ & $9.1$ & $10.3$ & $11.6$ & $13.2$\\
\hline
\hline
\end{tabular}
\label{rho_cs_w_tab}
\end{table}

\section{Conclusions}
\label{sect_5}
In this paper we have analyzed the impact of the dynamical scalar field quintessence dark energy on the NS. The density distribution was found from the numerical solutions of the conservation equation for NS matter and dark energy in joint potential, which corresponds to the static equilibrium of NS. We studied how this distribution depends on the parameters of dark energy (EoS $w_\infty$ and squared effective speed of sound $c_s^2$) and also, on the central density of baryonic matter $\rho_c$. We have found, that the relative deviation of the dark energy density $\delta_{de}(r)$ inside a neutron star increases as $w_\infty$ and $\rho_c$ grow and as $c_s$ decreases. 

We have also established, that there is a turn-off point in the dependence of NS mass on its central matter density determined by the amount of dark energy inside a star.
When exceeding this turn-off point, the dark energy with certain set of parameters ($c_s^2$ and $w_{\infty}$) makes impossible a stable static solution of equilibrium equation to exist, meaning that considered combination of parameters ($M$ and $\rho_c$ or $M-R$ relation) is impossible.

Using this, we have established, that current limitation on the maximal mass of NSs allows one to constrain the minimal value of speed of sound $c_s$ relying on certain NS model. We have used the Brussels-Montreal EoS for NS matter with the set of parameters BSk20-21 and the estimations of maximal NS mass from binary NS merger GW170817 to obtain the lowest possible minimum value for $c_s^2\gtrsim 10^{-2}$. We have also found that the dependence of the turn-off point on the background EoS parameter of dark energy $w_\infty$ is weak, so we didn't use it to establish constraints on $w_\infty$.

This constraint is stronger than one obtained from consideration of "white dwarfs + dark energy" system ($c_s^2\gtrsim  10^{-4}$), which means that NSs are more suitable objects for studying this dark component of the Universe. On the other hand, large uncertainties in the determination of masses of NSs, as well as knowledge of the matter state inside them, postpone obtaining interesting results for further perspective.

\section*{Acknowledgements}
This work was supported by the projects of Ministry of Education and Science of Ukraine $\Phi\Phi$-63Hp (No. 0117U007190) and “Formation and characteristics of elements of the structure of the multicomponent Universe, gamma radiation of supernova remnants and observations of variable stars” (No. 0119U002210).


\begin{thebibliography}{}

\bibitem {Creminelli2010}P.~Creminelli, G.~D'Amico, J.~Norena, L.~Senatore, F.~Vernizzi, 
\textit{Spherical collapse in quintessence models with zero speed of sound}, J. Cosmol. Astropart. Phys. \textbf{03}, 027 (2010).

\bibitem{Pace10} F.~Pace, J.-C.~Waizmann,  M.~Bartelmann, \textit{Spherical collapse model in dark energy cosmologies}, MNRAS \textbf{406}, 1865 (2010).

\bibitem{Tsizh2015} M.~Tsizh, B.~Novosyadlyj, \textit{Dynamics of dark energy in collapsing halo of dark matter}, Advances in Astronomy and Space Physics {\bf 15}, 51 (2015).

\bibitem{novosyadlyj_2016} B. Novosyadlyj, M. Tsizh, Yu. Kulinich, \textit{Dynamics of minimally coupled dark energy in spherical halos of dark matter}, General Relativity and Gravitation {\bf 48}, 30 (2016).

\bibitem{Babichev2013} E.~Babichev, V.~Dokuchaev, and Yu. Eroshenko, \textit{Black holes in the presence of dark energy, Usp. Fiz. Nauk} {\bf 183} (2013), 1287.

\bibitem{novosyadlyj_2014} B.~Novosyadlyj, Yu.~Kulinich, M.~Tsizh, \textit{Dynamics of dark energy in the gravitational fields of matter inhomogeneities, Phys. Rev. D} {\bf 90} (2014), 063004.

\bibitem{smerechynskyi19} S.~Smerechynskyi, M.~Tsizh, B.~Novosyadlyj, \textit{White dwarfs as a probe of dark energy, Phys. Rev. D} \textbf{101} (2020), 023001.

\bibitem{tsizh14} M.~Tsizh, B.~Novosyadlyj, Yu.~Kulinich, \textit{Distribution of dark energy in the vicinity of compact objects}, \textit{WDS'14 Proceedings of Contributed Papers--Physics} (MATFYZ-PRESS, Prague, 2014), 21–25. 

\bibitem{Kamiab11} F. Kamiab, N. Afshordi, \textit{Neutron Stars and the Cosmological Constant Problem, Phys. Rev. D} \textbf{84} (2011), 063011.

\bibitem{Alavirad13} H.~Alavirad, J. M.~Weller, \textit{Modified gravity with logarithmic curvature corrections and the structure of relativistic stars, Phys. Rev. D} \textbf{88} (2013), 124034.

\bibitem{Pani11} P.~Pani, E.~Berti, V.~Cardoso, J.~Read, \textit{Compact stars in alternative theories of gravity. Einstein-Dilaton-Gauss-Bonnet gravity, Phys. Rev. D} \textbf{84} (2011), 104035.

\bibitem{Yazadjiev15} S. S.~Yazadjiev, D. D.~Doneva, K. D.~Kokkotas, \textit{Rapidly rotating neutron stars in R-squared gravity, Phys. Rev. D}  \textbf{91} (2015), 084018. 

\bibitem{Sbisa20} F. ~Sbisa, P. O. Baqui, T. Miranda, S. E. Jorás, O. F. Piattella, \textit{Neutron star masses in $R^2$-gravity, Physics of the Dark Universe} \textbf{27} (2020), 100411.

\bibitem{Kpadonou15} A. V.~Kpadonou, M. J. S.~Houndjo, M. E.~Rodrigues, \textit{Tolman-Oppenheimer-Volkoff Equations and their implications or the structures of relativistic Stars in f(T) gravity, Astrophys. Space Sci.} \textbf{361} (2016), 244. 

\bibitem{Lavallaz10} A.~de Lavallaz, M.~Fairbairn, \textit{Neutron Stars as Dark Matter Probes, Phys. Rev. D} \textbf{81} (2010), 123521.

\bibitem{Guver13} T.~Guver, A.~Erkoca, M. H.~Reno, I.~Sarcevic, \textit{On the capture of dark matter by neutron stars, Journal of Cosmology and Astroparticle Physics} \textbf{5} (2014), 013.

\bibitem{Ellis18} J.~Ellis, G.~Hutsi, K.~Kannike, L.~Marzola, M.~Raidal, V.~Vaskonen, \textit{Dark Matter Effects On Neutron Star Properties, Phys. Rev. D} \textbf{97} (2018), 123007.

\bibitem{Ivanytskyi20}  O.~Ivanytskyi, V.~Sagun, I.~Lopes, \textit{Neutron stars: new constraints on asymmetric dark matter}, eprint arXiv:1910.09925 (2019).

\bibitem{Deliyergiyev19} M.~Deliyergiyev, A.~Del Popolo, L.~Tolos, M.~Le Delliou, X.~Lee, F.~Burgio \textit{Dark compact objects: an extensive overview, Phys. Rev. D} \textbf{99} (2019), 063015.

\bibitem{haensel_book_2007} P. Haensel, A.Y. Potekhin, and D.G. Yakovlev, \textit{Neutron Stars 1.Equation of State and Structure} (Springer-Verlag, New York, 2007).

\bibitem{Tolman} R.C. Tolman, \textit{Static Solutions of Einstein's Field Equations for Spheres of Fluid, Phys. Rev.} \textbf{55} (1939), 364.

\bibitem{Oppenheimer} J. R. Oppenheimer, G. M. Volkoff, \textit{On Massive Neutron Cores, Phys. Rev.} \textbf{55} (1939), 374.

\bibitem{Chandrasekhar_1931a} S. Chandrasekhar, \textit{The Maximum Mass of Ideal White Dwarfs, Astrophys. J.} \textbf{74} (1931), 81.

\bibitem{Chandrasekhar_1931b} S. Chandrasekhar, \textit{The Highly Collapsed Configurations of a Stellar Mass}, MNRAS \textbf{91}, 456 (1931).

\bibitem{Chandrasekhar_1935} S. Chandrasekhar, \textit{The Highly Collapsed Configurations of a Stellar Mass (second paper), MNRAS} \textbf{95} (1935), 207.

\bibitem{bombaci_1996} I. Bombaci, \textit{The maximum mass of a neutron star, Astronomy and Astrophysics} \textbf{305} (1996), 871.

\bibitem{alsing_2018} J. Alsing, H. O. Silva, E. Berti, \textit{Evidence for a maximum mass cut-off in the neutron star mass distribution and constraints on the equation of state, MNRAS} \textbf{478} (2018), 1377.

\bibitem{zhou_2019}  Y. Zhou, L. Chen, \textit{Ruling Out the Supersoft High-density Symmetry Energy from the Discovery of PSR J0740+6620 with Mass ${2.14}_{-0.09}^{+0.10}{M}_{\odot}$, Astrophys. J.} \textbf{886} (2019), 52.

\bibitem{annala_2018} E. Annala, T. Gorda, A. Kurkela, and A. Vuorinen, \textit{Gravitational-wave constraints on the neutron-star-matter Equation of State, Phys. Rev. Let.} \textbf{120} (2018), 172703.

\bibitem{zhang_2018} N. Zhang, B. Li, and J. Xu, \textit{Combined Constraints on the Equation of State of Dense Neutron-Rich Matter from Terrestrial Nuclear Experiments and Observations of Neutron Stars, Astrophys. J.} \textbf{859} (2018), 90.

\bibitem{zhang_2019} N. Zhang, B. Li, \textit{Implications of the mass $M=2.17^{+0.11}_{-0.10} M_{sun}$ of PSRJ0740+6620 on the Equation of State of Super-Dense Neutron-Rich Nuclear Matter, Astrophys. J.} \textbf{879} (2019), 99.

\bibitem{goriely_2010} S. Goriely, N. Chamel, J. M. Pearson, \textit{Further explorations of Skyrme-Hartree-Fock-Bogoliubov mass formulas. XII. Stiffness and stability of neutron-star matter, Phys. Rev. C} \textbf{82} (2010), 035804.

\bibitem{pearson_2011} J. M. Pearson, S. Goriely, N. Chamel, \textit{Properties of the outer crust of neutron stars from Hartree-Fock-Bogoliubov mass models, Phys. Rev. C} \textbf{83} (2011), 065810.

\bibitem{pearson_2012} J. M. Pearson, N. Chamel, S. Goriely, C. Ducoin, \textit{Inner crust of neutron stars with mass-fitted Skyrme functionals, Phys. Rev. C} \textbf{85} (2012), 065803.

\bibitem{potekhin_2013} A. Y.~Potekhin, A. F.~Fantina, N.~Chamel, J. M.~Pearson, S.~Goriely, \textit{Analytical representations of unified equations of state for neutron-star matter, Astron. Astrophys.} \textbf{560} (2013), A48.

\bibitem{cromartie_2019} H. T. Cromartie et al., \textit{Relativistic Shapiro delay measurements of an extremely massive millisecond pulsar, Nature Astronomy} \textbf{4} (2020), 72.

\bibitem{linares_2018} M. Linares, T. Shahbaz, and J. Casares, \textit{Peering into the Dark Side: Magnesium Lines Establish a Massive Neutron Star in PSR J2215+5135, Astrophys. J.} \textbf{859} (2018), 54. 

\bibitem{linares_2019} M. Linares, \textit{Super-Massive Neutron Stars and Compact Binary Millisecond Pulsars}, eprint arXiv:1910.09572 (2019).

\bibitem{kerkwijk2010} M. H. van Kerkwijk, R. Breton, S. R. Kulkarni, \textit{Evidence for a Massive Neutron Star from a Radial-Velocity Study of the Companion to the Black Widow Pulsar PSR B1957+20, Astrophys. J.} \textbf{728} (2011), 95.

\bibitem{romani_2015} R. W. Romani, A. V. Filippenko, and S. B. Cenko, \textit{A Spectroscopic Study of the Extreme Black Widow PSR J1311-3430, Astrophys. J.} \textbf{804} (2015), 115.

\bibitem{horvath_2017} J. E. Horvath and R. Valentim, \textit{The Masses of Neutron Stars}, in \textit{Handbook of Supernovae} (Springer International Publishing AG, 2017), p. 1317.

\bibitem{luo_2020} J. Luo, U. Pen,  M. Van Kerkwijk, P. Demorest, R. Main, \textit{Is the Black Widow Pulsar Eclipsed by its Companion?}, \textit{American Astronomical Society meeting} \textbf{235} (2020), id. 408.07. 

\bibitem{shibata_2019} M. Shibata, E. Zhou, K. Kiuchi, S. Fujibayashi, \textit{Constraint on the maximum mass of neutron stars using GW170817 event, Phys. Rev. D} {\textbf 100} (2019), 023015.

\bibitem{lawrence_2015} S. Lawrence, J. G. Tervala, P. F. Bedaque, M. C. Miller, \textit{An Upper Bound on Neutron Star Masses from Models of Short Gamma-Ray Bursts, Astrophys. J.} \textbf{808} (2015), 186.

\bibitem{fryer_2015} C. L. Fryer, K. Belczynski, E. Ramirez-Ruiz, S. Rosswog, G. Shen,  A. W. Steiner, \textit{The Fate of the Compact Remnant in Neutron Star Mergers, Astrophys. J.} \textbf{812} (2015), 24.

\bibitem{margalit_2017} B. Margalit, B. D. Metzger, \textit{Constraining the Maximum Mass of Neutron Stars from Multi-messenger Observations of GW170817, Astroph. J. Let.} \textbf{850} (2017), L19.

\bibitem{rezzolla_2018} L. Rezzolla, E. R. Most, L. R. Weih, \textit{Using Gravitational-wave Observations and Quasi-universal Relations to Constrain the Maximum Mass of Neutron Stars, Astrophys. J. Let.} \textbf{852} (2018), L25.

\bibitem{ruiz_2018} M. Ruiz, S. L. Shapiro, A. Tsokaros, \textit{GW170817, general relativistic magnetohydrodynamic simulations, and the neutron star maximum mass, Phys. Rev. D} \textbf{97} (2018), 021501(R).

\bibitem{abbott_2017} B. P. Abbott et al., \textit{GW170817: Observation of Gravitational Waves from a Binary Neutron Star Inspiral, Phys. Rev. Let.} \textbf{119} (2017), 161101.

\bibitem{novosyadlyj_2013}B.~Novosyadlyj, M.~Tsizh, Yu.~Kulinich, \textit{Gravitational stability of dark energy in galaxies and clusters of galaxies,	Kinematics and Physics of Celestial Bodies} \textbf{30} (2014), 53.

\bibitem{armendariz-picon_1999} C. Armendariz-Picon, T. Damour, and V. Mukhanov, \textit{k-inflation, Phys. Let. B} \textbf{458} (1999), 209.

\bibitem{Sergijenko14} O.~Sergijenko, B.~Novosyadlyj, \textit{Sound speed of scalar field dark energy: weak effects and large uncertainties, Phys. Rev. D} {\bf 91} (2015), 083007.
\end{thebibliography}
\end{document}